# HD 100453: A LINK BETWEEN GAS-RICH
# PROTOPLANETARY DISKS AND GAS-POOR DEBRIS DISKS

K. A. Collins[1,2], C. A. Grady[3], K. Hamaguchi[4,5], J. P. Wisniewski[6,7], S. Brittain[8], M. Sitko[9], W. J. Carpenter[10],
J. P. Williams[11], G. S. Mathews[11], G. M. Williger[1,12,13], R. van Boekel[14], A. Carmona[15,14,16], Th. Henning[14],
M. E. van den Ancker[16], G. Meeus[17], X. P. Chen[14], R. Petre[18], B. E. Woodgate[18]

## ABSTRACT

HD 100453 has an IR spectral energy distribution (SED) which can be fit with a power-law plus a blackbody. Previous analysis of the SED suggests that the system is a young Herbig Ae star with a gas-rich, flared disk. We reexamine the evolutionary state of the HD 100453 system by refining its age (based on a candidate low-mass companion) and by examining limits on the disk extent, mass accretion rate, and gas content of the disk environment. We confirm that HD 100453B is a common proper motion companion to HD 100453A, with a spectral type of M4.0V − M4.5V, and derive an age of $10 \pm 2$ Myr. We find no evidence of mass accretion onto the star. Chandra ACIS-S imagery shows that the Herbig Ae star has $L_x/L_{bol}$ and an X-ray spectrum similar to non-accreting β Pic Moving Group early F stars. Moreover, the disk lacks the conspicuous Fe II emission and excess FUV continuum seen in spectra of actively accreting Herbig Ae stars, and from the FUV continuum, we find the accretion rate is $< 1.4 \times 10^{-9}$ M$_\odot$ yr$^{-1}$. A sensitive upper limit to the CO J = 3−2 intensity indicates that the gas in the outer disk is likely optically thin. Assuming a [CO]/[H$_2$] abundance of $1 \times 10^{-4}$ and a depletion factor of $10^3$, we find that the mass of cold molecular gas is less than ~0.33 M$_J$ and that the gas-to-dust ratio is no more than ~4:1 in the outer disk. The *combination* of a high fractional IR excess luminosity, a relatively old age, an absence of accretion signatures, and an absence of detectable circumstellar molecular gas suggests that the HD 100453 system is in an unusual state of evolution between a gas-rich protoplanetary disk and a gas-poor debris disk.

*Subject headings:* binaries: visual — planetary systems: protoplanetary disks — stars: individual (HD 100453, 51 Eri, SAO 206462, DX Cha, HD 163296, HD 169142, HD 100546, 2MASSWJ1207334−393254) — stars: low-mass, brown dwarfs — stars: pre-main-sequence

[1] Department of Physics and Astronomy, University of Louisville, Louisville, KY 40292, USA; karen.collins@insightbb.com, williger@physics.louisville.edu

[2] Kentucky Space Grant Consortium Fellow

[3] Eureka Scientific and GSFC, 2452 Delmer, Suite 100, Oakland CA 96002, USA; Carol.A.Grady@nasa.gov

[4] CRESST and X-ray Astrophysics Laboratory NASA/GSFC, Greenbelt, MD 20771; Kenji.Hamaguchi@nasa.gov

[5] Department of Physics, University of Maryland, Baltimore County, 1000 Hilltop Circle, Baltimore, MD 21250

[6] Department of Astronomy, University of Washington, Box 351580 Seattle, WA 98195; wisniewski@astro.washington.edu

[7] NSF Astronomy & Astrophysics Postdoctoral Fellow

[8] Clemson University, Clemson, South Carolina, 29634; sbritt@clemson.edu

[9] Space Science Institute, University of Cincinnati; Michael.Sitko@uc.edu

[10] University of Cincinnati; wcarpenter4@cinci.rr.com

[11] Institute for Astronomy, University of Hawaii, 2680 Woodlawn Dr., Honolulu, HI 96822; jpw@ifa.hawaii.edu, gmathews@ifa.hawaii.edu

[12] Department of Physics & Astronomy, Johns Hopkins University

[13] Department of Physics, Catholic University

[14] Max-Planck-Institut für Astronomie, Königstuhl 17, 69117 Heidelberg, Germany; boekel@mpia-hd.mpg.de, henning@mpia-hd.mpg.de, chen@mpia-hd.mpg.de

[15] ISDC Ch. d'Ecogia 16, CH-1290 Versoix, Switzerland & Observatoire de Genève, University of Geneva, Ch. des Maillettes 51, 1290 Versoix, Switzerland; andres.carmona@unige.ch

[16] European Southern Observatory, Karl Schwarzschild Strasse 2, 85748 Garching bei München, Germany; mvandena@eso.org

[17] Astrophysikalisches Institut Potsdam (AIP), An der Sternwarte 16, 14482 Potsdam, Germany; gwen@aip.de

[18] NASA Goddard Space Flight Center, Greenbelt, MD 20771, USA; petre@lheavx.gsfc.nasa.gov, Bruce.E.Woodgate@nasa.gov

## 1. INTRODUCTION

Observational investigations of our solar system and theoretical studies indicate that gas-giant planets form in less than ~10 Myr and Earth-like terrestrial planets in less than ~30 Myr (Zuckerman & Song 2004). Pre-main sequence (PMS) A-type stars provide an ideal laboratory to study the early stages in planetary system evolution. They are of high enough luminosity that associated disks have been spatially resolved out to ~400 pc[19], are cool enough that they are unlikely to create infrared excesses in the form of ionized gas (see Su et al. 2006), and can retain heavier and more extended disks than lower mass stars and thus might form massive planets at wider separations. Radial velocity studies show an excess of Jovian mass bodies around main sequence A-type stars (Johnson et al. 2007), and the only confirmed direct imaging detections of exoplanets (hosted by a stellar mass object) have occurred with A-type stars (Marois et al. 2008; Kalas et al. 2008).

Observations of nearby, *isolated* A-type stars (i.e. stars that are no longer associated with their parent molecular cloud) benefit from relatively low levels of interstellar extinction and field contamination. They are far from the extreme environments of H II regions near OB associations, and thus, the stars themselves, and not their environments, drive their disk structure, chemistry, and evolution. As a result, the characteristics of isolated Herbig Ae stars in the age range of ~10 Myr are of particular interest, *if they can be securely dated*. Reliable dating of isolated A stars is difficult after about 5 Myr (Stauffer 2004) due to the rapid evolution of the star to the Zero-Age Main Sequence (ZAMS). However, the identification of co-moving low-mass stars which are associated with A stars (e.g. Zuckerman & Song 2004) permits independent dating of the A star system. With a reliable age, Herbig Ae disk environments in which the gas has dissipated place constraints on the time available for gas-giant planet formation and migration.

HD 100453A is a well studied, isolated A9Ve Herbig Ae star (Houk & Cowley 1975) identified by Chen et al. (2006) as a candidate binary with a suspected co-moving early-M companion (also referred to hereafter as HD 100453B). The A star has a mass of 1.7 $M_\odot$ (Dominik et al. 2003) and a distance of $111^{+11}_{-8}$ pc (Perryman et al. 1997). By placing the A-type star on PMS evolutionary tracks, Meeus et al. (2002) (hereafter M02) suggest that HD 100453A is close to the ZAMS with t $\geq$ 10 Myr. HD 100453A appears to be a member of the Lower Centaurus Crux Association (Kouwenhoven et al. 2005), which suggests t $\leq$ ~20 Myr (Mamajek et al. 2002). Garcia Lopez et al. (2006) find a mass accretion rate of $9\times10^{-9}$ $M_\odot$ yr$^{-1}$ based on the equivalent width (EW) of the Brγ emission line. HD 100453 shows an absence of the 10 μm silicate feature but displays PAH emission (Meeus et al. 2001, hereafter M01; van Boekel et al. 2005) and strong indications of cold crystalline silicate (forsterite) emission at 34 μm (Vandenbussche et al. 2004).

The combination suggests grain growth in the inner regions of the disk.

HD 100453 has an IR spectral energy distribution (SED) which can be fit with a power-law plus a blackbody (M01). Herbig Ae/Be stars showing these SED characteristics are classified as Meeus group I sources and are interpreted as arising from gas-rich, flared disks where small grains are well mixed with the gas (M01; Dullemond & Dominik 2004a,b). M01 suggested that the group I SEDs are precursors to SEDs which can be fit with a power-law alone (i.e. Meeus group II objects). More recently, the disks of some Meeus group I sources have been spatially resolved (e.g. HD 100546, Grady et al. 2005; HD 135344B, Brown 2007, Pontoppidan et al. 2008) and show inner disk dust holes (i.e. transitional disks; Najita et al. 2007) or gaps (i.e. pre-transitional disks; Espaillat et al. 2007), suggesting that a subset of Meeus group I disks are in an intermediate state of evolution between primordial disks and debris disks. For the pre-transitional and transitional disks, as for the flared disks, the assumption is that the outer portion of the disk is rich in molecular gas. Except for a marginal detection of spatially resolved polycyclic aromatic hydrocarbons (PAHs) (Habart et al. 2006), the disk of HD 100453 has not been spatially resolved to date.

In this work, we utilize new multi-wavelength observations to further investigate the evolutionary state of the HD 100453 system by refining the age of the system based on a candidate co-moving, low-mass companion and by examining limits on the size of the disk, the mass accretion rate, and gas content of the disk environment. To test the binary hypothesis, we utilize second epoch narrowband NIR imagery to investigate common proper motion, NIR spectroscopy to determine the spectral type, and X-ray data to test the T Tauri nature of the companion candidate. With the spectral type and Chen et al. (2006) photometry, we locate the companion candidate on PMS tracks to estimate its age and mass. We examine indicators of mass accretion rate as well as cold, warm, and hot gas tracers to characterize the gas content of the A star disk environment. We also examine high resolution direct imagery and point spread function (psf) subtracted coronagraphic imagery to place limits on the disk extents.

## 2. OBSERVATIONS AND DATA REDUCTION

### 2.1 *Chandra ACIS-S Imaging Spectroscopy*

HD 100453 was observed by the *Chandra* X-Ray Observatory (program P07200493; sequence 200403; Grady, PI) on 2006 March 10 for 9.7 ksec using the Advanced CCD Imaging Spectrometer (ACIS: Garmire et al. 2003). To ensure the highest soft energy response, HD 100453 was placed at the ACIS-S aimpoint, and data were obtained using the default 8.3′ × 50.6′ field of view. When used as an imager in this configuration, ACIS-S has a resolution of 0″.5 (FWHM) and obtains pulse height spectra from $0.3 - 8.0$ keV.

---

For analysis, we used the CIAO[20] software package ver. 3.4 (CALDB ver. 3.3.0.1) and generated response matrices and auxiliary files using `psextract` and `mkacisrmf`. To improve the spatial resolution, we utilized the sub-pixel event repositioning (SER) algorithms (Tsunemi et al. 2001; Li et al. 2003, 2004).

For the absolute position determination of the *Chandra* image frame, we cross-correlated the X-ray sources detected on the same backside-CCD chip (ID=7) with the Point Source Catalog of the 2MASS All-Sky Data Release[21]. Among 8 X-ray sources other than HD 100453 detected with the *wavedetect* package at above the 4σ significance, 2 sources have 2MASS counterparts. The average offset of the *Chandra* image is less than 0″.1 from the 2MASS frame without any positional correction.

We generated a sky coordinate image with a fine spatial resolution of 0″.125 pixel$^{-1}$ (Fig. 1a). There are two peaks, each of which corresponds to the position of HD 100453A (NW peak) and to the companion candidate, HD 100453B (SE peak), respectively. The color-coded image (Fig. 1b) clearly shows that HD 100453B has harder emission. The image also shows apparent tail structures around either source toward the NW (HD 100453A) and NE-SW (HD 100453B) directions. The psf itself does not have such a structure, but we cannot confidently say if the structures are real, considering unknown factors in the image reconstruction process and limited photon statistics.

We extracted X-ray events within r = 0″.6 radius from each peak (Fig. 1a), which includes ~81% of the total flux from a point source based on a psf image at 0.8 keV generated with `mkpsf`. A total of 51 counts were extracted for HD 100453A and a total of 52 counts were extracted for HD 100453B in the 0.3 − 8 keV energy range. The data were binned into 2 ksec bins, and neither source showed significant time variation during the observation, but given the low count rate, the limit is not stringent. We extracted an X-ray spectrum of each source from the same source region and subtracted background obtained from a surrounding source free region. Neither of the spectra can constrain the H column density ($N_H$) due to the limited photon statistics. Since HD 100453A has small optical extinction, E(B − V) < 0.08 (Malfait et al. 1998), we assume the same for HD 100453B and adopt a thin-thermal plasma model for both (APEC[22] code), without X-ray absorption in the spectral fits. The fitting result and spectra are shown in Table 1 and Fig. 1c,d, respectively. Both sources fit as single temperature, unabsorbed models. The intrinsic X-ray flux increases by ~40% for $kT$~0.3 keV plasma between 0.35 − 2 keV if $N_H = 5.4 \times 10^{20}$ cm$^{-2}$, equivalent to E(B − V) = 0.08 (Ryter 1996). We were also able to fit single temperature, absorbed models with E(B − V) < 0.75 for the primary star and E(B − V) < 1.24 for the companion candidate, with comparable reduced $\chi^2$ results. Either model fits satisfactorily.

## 2.2 *Imagery*

### 2.2.1 *HST ACS/SBC Imagery*

HD 100453 was observed with the Hubble Space Telescope (HST) and the Advanced Camera for Surveys (ACS) Solar Blind Channel (SBC) in filter F122M (nominal λ = 1210 Å, FWHM = 60 Å) on 2006 June 7 (HST-GO-10764; Grady, PI). The HD 100453 data, with total integration time of 2096 s, were obtained in 4 exposures within a single spacecraft orbit at a spacecraft orientation of −82.0755° (Table 2). The known single source white dwarf star HS 2027+0651 was also observed with the SBC in filter F122M to serve as a psf reference star (Table 2).

For each exposure, the field was dithered using the ACS small-box dither pattern (Gonzaga et al. 2005, 2006, with dither offsets of 0″.15 − 0″.19). The data were reduced following Pavlovsky et al. (2006). In the final, geometrically corrected imagery (Fig. 2a), the pixel scale is 0″.025 × 0″.025 and the background level is 1.1×10$^{-3}$ counts pixel$^{-1}$ s$^{-1}$ with a standard deviation of 1.3×10$^{-3}$ counts pixel$^{-1}$ s$^{-1}$. There are no sources detectable at the 5σ level or higher except for the brightest source in the field, HD 100453A, and the recently identified optical ghost to the upper right of HD 100453A (Collins et al. 2007).

### 2.2.2 *HST ACS/HRC Imagery*

HD 100453 was observed using the direct and coronagraphic modes of the ACS High Resolution Channel (HRC) camera on 2003 November 12 (GTO program 9987). The F606W filter imaging was obtained using the ACS 1″.8 diameter occulting spot, yielding a 29″.2 × 26″.2 field of view with a pixel scale of 0″.028 × 0″.025 (Table 2). HD 87427 (F0V), a star of similar spectral type to HD 100453 (A9V), was observed immediately following HD 100453 to serve as a psf reference star (Table 2). The data have been reduced following Wisniewski et al. (2008) yielding an image with 0″.025 × 0″.025 size pixels.

The resulting 7″.5 × 7″.5 direct F606W image of HD 100453 is shown in Fig. 2b. While the central HD 100453A source is clearly saturated, a distinct second source is clearly visible ~1″.05 southeast from HD 100453A, at a position angle of 126°. Fig. 2c is a 15″.0 × 15″.0 psf-subtracted coronagraphic image of HD 100453 in the F606W filter, plotted on a linear scale and smoothed using a 3 × 3 Gaussian kernel. No evidence of an extended scattered light disk is visible at r > 250 AU. HD 100453B, identified within the green circle, is clearly detected through the psf-subtraction residuals (see inset radial profile) and consistent with the position determined from the direct imaging.

SYNPHOT, the synthetic photometry package within STSDAS, was used to determine the correction factor needed to calibrate our data to an absolute photometric scale. We used a synthetic spectrum of HD 508 (A9IV), normalized to a V-band magnitude of 7.79, as a template for HD 100453A, and a synthetic spectrum of χ Ser (F0IV), normalized to a V-band magnitude of 5.72, as a template for HD 87427. All psf-subtracted, distortion corrected images were normalized to the synthetic flux of HD 100453A. Note that all photometry reported is based on the STMAG photometric system, and photometry extracted from the coronagraphic data have been corrected for the known 52.5% reduction of flux induced by the occulting spot (see § 3.3 for photometry results).

### 2.2.3 VLT NACO Imagery

HD 100453 was observed with NACO (Lenzen et al. 2003; Lagrange et al. 2003) during the nights of 2006 May 23 and 2006 June 22 (Table 3). NACO is composed of NAOS, the first Adaptive Optics System on the VLT (Rousset et al. 2003) and CONICA, a $1 - 5$ μm imaging, coronagraphic, spectroscopic and polarimetric instrument (Lenzen et al. 1998). HD 100453 was previously observed with NACO during the night of 2003 June 2, through the Brγ narrow band filter (Chen et al. 2006). A total exposure time of 3.5 seconds was used, at an airmass of 1.15 and an ambient seeing of 0″.9.

For the Brγ data, the S27 camera was used, which provides a pixel scale of 27.15 mas pixel$^{-1}$ and a field of view of about $28'' \times 28''$. Standard pipeline processing was applied, consisting of dark subtraction and flat fielding. For the L′−band and M′−band data, the coronagraphic mode was used with chopping and nodding to reduce the background contamination. In addition, a pupil stop was used for the science observations to reduce stray light. A calibration star was observed in the L′−band and M′−band to facilitate photometry, but without the pupil stop. All reported photometry (§ 3.3) has been corrected for the known 0.234 mag reduction of flux induced by the pupil stop.

### 2.3 Spectroscopy

#### 2.3.1 FUSE Spectroscopy

HD 100453 was observed by FUSE on 2003 June 10 using the large aperture (30"×30") in time-tag mode (C1260101; Grady PI). For comparison spectra, we use data for DX Cha (HD 104237A, P2630101 from 2001 May 25; Wilkinson, PI ) and Altair (P1180701; Linsky, PI). DX Cha has a UV spectrum intermediate between A7V and A8V (Grady et al. 2004), has variable UV excess light, and drives a bipolar jet, while Altair is a nearby A7V, older main sequence star. The DX Cha data have previously been discussed in Grady et al. (2004), while the Altair data were discussed by Simon et al. (2002).

Data for all three observations were reduced using CALFUSE 3.0.8 and stacked independently by channel segment using the *corrcal* routine written by S. Friedman. The total exposure time was 11,788 s for HD 100453, 20,855 s for DX Cha, and 4234 s for Altair. However, only exposures that were useful or free of obvious defects were included in the summed data (Fig. 3). HD 100453 shows the FUV emission typical of an early F star.

When compared with the Altair spectrum scaled to the V magnitude of HD 100453, HD 100453 shows excess signal in C III λ1176 Å, indicating the presence of chromospheric activity. Its continuum in the region of the LiF2A detector (Fig. 4a) lies well above the level expected for the photosphere of an A7V star. We therefore conclude that the FUSE spectrum of HD 100453 has negligible photospheric contamination.

DX Cha and HD 100453 lie at essentially the same distance, to within the Hipparcos error bars. Adopting a spectral type for DX Cha of A7.5V, we find negligible selective extinction, a result similar to HD 100453. As a consequence, the HD 100453 and DX Cha spectra can be directly compared (Fig. 4b). The 2001 DX Cha spectrum, which is representative of 3 observations spanning 6 years, shows the FUV excess continuum light expected of a star which is actively accreting (Grady et al. 2004), as well as the double-peaked C III λ1176 Å emission profile seen in other jet-driving Herbig Ae stars (e.g. HD 163296 Devine et al. 2000 + Deleuil et al. 2005). In contrast, HD 100453 has the blocky, multi-component emission typical of chromospheric emission in moderately rotating late-type stars (see Redfield et al. 2002). Table 4 shows the mean of the continuum flux (plus some spectral features) in a 10 Å bandpass centered on 1160 Å for both DX Cha and HD 100453. Since the HD 100453 flux is clipped at less than $-6 \times 10^{-15}$ erg cm$^{-2}$ s$^{-1}$ Å$^{-1}$, we report an upper limit for its mean continuum. The 1σ upper limit to the HD 100453 mean continuum is a factor of 17 below the firm detection for DX Cha. Comparison of multi-epoch FUSE, IUE, and HST data, each covering different wavelength regimes, suggests that DX Cha's UV flux can vary by a factor of $3 - 4$. However, during the time period spanned by 4 epochs of DX Cha FUSE observations (2000 − 2006), we detect no significant FUV continuum variability in the range $1155 - 1165$ Å. As we can not definitively establish the state of DX Cha's FUV flux at the time of the HD 100453 FUSE observations, we can not quantify how the UV variability affects our DX Cha to HD 100453 FUV continuum ratio. We suggest that, at most, this uncertainty could influence the ratio by a factor of $3 - 4$, but given the consistency of the 4 epochs of DX Cha FUSE observations, we adopt the values listed in Table 4 to compare to the HD 100453 FUSE observation.

#### 2.3.2 IUE Archival Data

IUE data for HD 100453 are discussed in Malfait, Bogaert, and Waelkens (1998). The FUV spectrum, SWP 53937, shows Lyman alpha and other lines in emission, together with the photospheric spectrum expected for a late A-early F star (Fig. 5).

#### 2.3.3 VLT SINFONI Spectroscopy

HD 100453B was observed with the ESO-VLT Spectrograph for Integral Field Observations in the Near Infrared (SINFONI) mounted on the UT4 "Yepun" telescope on 2007 May 19 (Carmona, PI). SINFONI is a NIR ($1.1 - 2.45$ μm) integral field spectrograph fed by an adaptive optics module, installed at the Cassegrain focus of UT4 (Eisenhauer et al. 2003, Bonnet et al. 2004). The observations were performed with 3 gratings providing a spectral resolution around 2000, 3000, 4000 in the J, H and K band respectively (Table 5). We employed the spatial resolution of $0''.025$ per image slice, which corresponds to a field-of-view of $0''.8 \times 0''.8$. HD 100453A was used as a natural adaptive optics guide star, and no spectral dithering was used.

To perform the telluric correction, the spectrophotometric standard star HIP 057432 was observed immediately following the science observations employing an exposure time of 50, 40, and 30 seconds in the J, H, K bands, respectively. In each band, two offset positions were taken. At the end of the night, the spectrophotometric standard star HIP 051940 was observed in the J, H, and K bands employing an exposure time of 50, 40 and 20 seconds respectively. In each band, two offset positions were taken.

The raw frames of the target and standard stars were reduced using the SINFONI pipeline (Modigliani et al. 2007). From the final coadded wavelength calibrated datacubes of HD 100453B and the standard stars, the one-dimensional (1−D) spectra of HD 100453B were extracted. Each datacube was coadded in the wavelength direction producing a single two-dimensional (2−D) frame. We used a psf radius of 7 pixels and a background region size of 5 pixels for HD 100453B and sampled the FOV by a 2−D frame of 94 × 90 pixels in the J band, 100 × 96 pixels in the H band, and 84 × 80 pixels in the K band. For the standard stars, we used a psf radius of 15 pixels and a background region size of 10 pixels and sampled the FOV by a 2-D frame of 64 × 64 pixels in all three bands. The average number of counts per pixel in the background region was subtracted from each pixel in the psf region for each bandpass, and the total number of counts in the psf region was summed. Following this procedure, the 1-D spectrum was determined for HD 100453B and the standard stars in each of the observed bands.

Finally, to correct for telluric absorption, the 1−D extracted science spectrum of HD 100453B was divided by the 1−D extracted spectrum of each standard star. Small offsets of a fraction of a pixel in the wavelength direction were applied to the standard star spectrum until the best signal-to-noise ratio, in the corrected science spectra, was obtained. Since we observed two standard stars, two sets of telluric corrected spectra were obtained for HD 100453B in each band. We selected the spectra that exhibited the best signal-to-noise ratio. The spectra were normalized by fitting a second-degree polynomial to the continuum (Fig. 6).

### 2.3.4 Gemini South Phoenix Echelle Spectroscopy

The CO observations were taken with the Phoenix echelle spectrograph on Gemini South Observatory on 2007 February 6 with an exposure time of 480 seconds. Phoenix is a high-resolution $\lambda/\Delta\lambda = 50,000$ (4 pixel slit − $0''.34$), near-infrared $1 - 5$ μm spectrometer. An individual spectrum is single order and covers a very narrow wavelength range of 0.5% of the central wavelength, corresponding to a radial velocity range of 1500 km/sec (Hinkle et al. 1998; Hinkle et al. 2000; Hinkle et al. 2003). We observed HD 100453 with one M-band setting centered at 4.97 μm. The data were reduced and calibrated following DiSanti et al. (2001), Brittain et al. (2003), and Brittain (2004). The spectrum was divided by the spectrum of a standard star, HR 5671, to correct for the telluric absorption. The gaps in the spectrum are of areas with less than 50% transmittance. The CO emission lines P30 and P31 were not detected above noise in our observations (Fig. 7).

### 2.3.5 JCMT HARP Spectroscopy

We searched for the 345 GHz (868 μm) CO J = 3−2 line using the Heterodyne Array Receiver Program (HARP) (Dent et al. 2000) at the James Clerk Maxwell Telescope (JCMT) at Mauna Kea, Hawaii, on 2008 February 22 and 24. The weather was stable and dry with a precipitable water vapor level less than 2 mm. The observations were carried out in beam-switching mode with a chop of $120''$. The total on-source integration time was 20 minutes. Calibration was checked via similar observations of IRC+10216 and is estimated to be accurate within 5%. The rms radiation temperature in binned 10 km s$^{-1}$ channels was 0.0077 K, and the CO J = 3−2 emission line was not detected above noise.

### 3.0 RESULTS

#### 3.1 Relative Proper Motion of Objects in the Field

Analysis of both the direct (Fig. 2b) and coronagraphic (Fig. 2c) HST/ACS F606W images of HD 100453A reveals clear evidence of the companion candidate located at $1''.045 \pm 0''.025$ from the central star at a position angle of $126 \pm 1°$ measured E of N. In the direct image, we also detect a faint third object in the field at ∼$0''.8$ from HD 100453A at a position angle of ∼30° measured E of N. We detect a faint object near the same location in the VLT NACO J-band coronagraphic image. For clarity, we refer to this object as star C.

The proper motion (PM) of the HD 100453 system is $\mu_\alpha = -36.95 \pm 0.78$ mas yr$^{-1}$ and $\mu_\delta = -4.72 \pm 0.53$ mas yr$^{-1}$ (Perryman et al. 1997). We measured the relative position of HD 100453A and the companion candidate (HD 100453B) in the 2003 and 2006 epoch VLT NACO imagery by fitting 2−D Gaussians to the Brγ narrow band images at the approximate positions of each star. Comparing the positions for these two epochs (Table 6), we find a relative motion of $20.11 \pm 5.42$ mas at $238.40 \pm 28$ degrees E of N or $18.33 \pm 6.39$ mas in the direction of system PM. The relative motion of the companion candidate with respect to the primary star is $16.13 \pm 5.63$% of the system PM. Thus, we show that HD 100453A and HD 100453B are common proper motion companions.

To estimate the relative proper motion of star C compared to HD 100453A, the location of the primary star is assumed to be at the intersection of the diffraction spikes in both epochs of imagery. It is important to note that the optical axis of NACO does not coincide with the optical axis of the Nasmyth focus which could lead to inaccuracies in defining the location of the star under the coronagraph (D. Dobrzycka, private communication). Fig. 8 is an overlay of the NACO image onto the ACS image with the centers of the ACS and NACO diffraction spikes and object locations marked in red and blue, respectively. The *relative* motion of star C with respect to HD 100453A has both an amplitude and direction opposite to HD 100453A's proper motion, indicating that C is likely a background star.

## 3.2 HD 100453B Spectral Type Determination

The spectral type of HD 100453B is determined by comparing our VLT SINFONI spectra with the IRTF NIR spectral library[23] (Rayner et al. in prep). In Fig. 6 we present the continuum normalized J, H, and K spectra of HD 100453B and the continuum normalized spectral templates of M3.5V to M5.0V stars from the IRTF NIR spectral library.

In the J−band, the absorption features between $1.16\,\mu m$ − $1.20\,\mu m$ and $1.25\,\mu m$ − $1.27\,\mu m$ are consistent with spectral types M3.5V − M4.5V. In the H−band, the doublet near $1.67\,\mu m$ is consistent with spectral types M3.5V − M4.0V and the absorption feature near $1.52\,\mu m$ is consistent with spectral types M3.5V − M4.5V. In the K−band, the triplet at $1.97\,\mu m$ and the doublet at $2.20\,\mu m$ are consistent with spectral types M4.5V − M5.0V. We therefore adopt a spectral type of M4.0 − M4.5V for HD 100453B.

## 3.3 HD 100453B Photometry

We extract crude photometry for the companion from the direct (Fig. 2b) and coronagraphic (Fig. 2c) HST ACS F606W images, using a small $0''.1$ aperture and aperture corrections estimated via *SYNPHOT* (Table 7). The dominant source of uncertainty in these values is contamination contributed by the saturated central source in the direct imaging data and the contamination from significant psf-subtraction residuals in the coronagraphic imaging. Given the range of derived photometry, we suggest an approximate $m_{F606W}$ magnitude for the companion of $15.7 \pm 0.2$. Aperture corrections suggested by Bohlin (2007) to compensate for imaging a very red star in the wide F606W bandpass are negligible (<0.03 mag) compared to our other sources of uncertainty.

We extract L−band photometry for HD 100453B from our NACO L′−band coronagraphy, using the transformation for a M4.0V star given by Bessell and Brett (1988) (Table 7). We also transformed the $K_s$−band photometry reported by Chen et al. (2006) to the K-band using relations provided by Persson et al. (1998) (Table 7). Assuming B − V = 1.65 (Leggett et al. 1992), we used the techniques outlined in Sirianni et al. (2005) to convert the ACS $m_{F606W}$ flux to the V-band. Comparison of the V−K and V−L colors of HD 100453B to those of an unreddened M4V star (Leggett et al. 1992), indicates $E(V − K) = 0.06 \pm 0.20$ and $E(V − L) = 0.13 \pm 0.20$.

These colors indicate that HD 100453B has no significant detectable IR excess in the K-band or L-band.

## 3.4 HD 100453B X-ray Luminosity

To determine $\log(L_x/L_{bol})$ for HD 100453B, we utilize the V−band magnitude of 15.88 and the bolometric correction for an M4.0 star of −2.45 (Cox et al. 2001) to arrive at a value of $L_{bol} = 1.66 \times 10^{32}$ erg s$^{-1}$. From our Chandra data for the companion (Table 1), we have $L_x = 3.3 \times 10^{28}$ erg s$^{-1}$ in the range 0.3 − 2.0 keV. The result is $\log(L_x/L_{bol}) = -3.70$, which indicates the companion is an active late-type star (Feigelson et al. 2003).

## 3.5 Dating the System

Chen et al. (2006) assumed an age of 10 − 20 Myr for HD 100453B and utilized a PMS Hertzsprung-Russell (H−R) diagram to find that the $K_s$−band magnitude is consistent with a 0.3 M$_\odot$ star, corresponding to a spectral type of M3 − M5. With a measured spectral type of M4.0V − M4.5V and photometry, we utilize PMS H−R diagrams to refine the age estimate of the HD 100453 system.

We use the Siess et al. (2000) PMS stellar model online "WWW tools"[24] to create PMS H−R diagrams containing isochrones and mass tracks with magnitude as a function of effective temperature ($T_{eff}$). The error bars in magnitude are dominated by uncertainties in the photometry and distance to the system. The error bars in $\log(T_{eff})$ are dominated by uncertainties in the spectral type and the mapping between spectral type and $T_{eff}$. We use the spectral type to $T_{eff}$ conversions in Kenyon and Hartmann (1995) for the primary star and Luhman et al. (2003) for the companion.

Adopting $T_{eff} = 3200 − 3300$ K for HD 100453B and using the Siess et al. (2000) H−R diagram (Fig. 9a), we find a mass of 0.16 − 0.21 M$_\odot$ and an age of 8 − 12 Myr. We also examine the PMS tracks of Baraffe et al. (1998) and find a mass of 0.17 − 0.24 M$_\odot$ and an age of 7 − 12 Myr.

As a consistency check, we examine the Siess et al. (2000) PMS H−R diagram of HD 100453A (Fig. 9b). We utilize V−band photometry to avoid contamination from IR excess and correct for reddening along the sight line. Using 51 Eri (F0V) as a reference (B − V = 0.28, Johnson et al. 1966) and a grid of colors from Cox et al. (2001), we calculate B − V = 0.27 for an A9V star. Based on HD 100453A observed magnitudes of B = 8.07 and V = 7.78 (Table 7), the color excess is $E(B − V) = 0.02$. Assuming R = 3.1, then $A_v = 0.06$ and $V_{Intrinsic} = 7.72$. Adopting $T_{eff} = 7300$ K − 7500 K, we find a mass of 1.65 − 1.82 M$_\odot$ and an age of 9 − 18 Myr. The Baraffe et al. (1998) tracks cannot be compared for HD 100453A since the models do not provide data for mass tracks >1.4 M$_\odot$.

The age estimates of the system based on both the Siess et al. (2000) and the Baraffe et al. (1998) models for the companion are in agreement with the Kouwenhoven et al. (2005) result that HD 100453A is a member of the Lower Centaurus Crux Association, which places an upper limit for the age at about 20 Myr (Mamajek et al. 2002). We conclude that the age of the HD 100453 system is $10 \pm 2$ Myr and that the mass of HD 100453B is $0.20 \pm 0.04$ M$_\odot$.

---

[23] http://irtfweb.ifa.hawaii.edu/ spex/spexlibrary/IRTFlibrary.html

[24] http://www-astro.ulb.ac.be/~siess/prog.html

### 3.6 HD 100453A Accretion

Accretion onto PMS stars produces not only enhanced emission in transitions otherwise associated with chromospheres and transition regions, but also an enhanced UV continuum flux which can be detected against the light of a late A star shortward of 2000 Å, plus jets and/or Herbig-Haro knots, which can be imaged in Lyα (Calvet et al. 2004; Grady et al. 2004). These emission characteristics are seen both for Meeus group I objects like AB Aur (Endres et al. 2005), and for Meeus group II objects like HD 163296 (Devine et al. 2000), HD 104237 (Grady et al. 2004), and MWC 480 (Stecklum et al. 2008).

#### 3.6.1 X-Ray Limits

From our Chandra data, $L_x$ in the range $0.3 - 2$ keV for HD 100453A is $4.0 \times 10^{28}$ erg s$^{-1}$ (Table 1) and is just below the low end of the range for actively accreting Herbig Ae stars of $L_x \geq 10^{29}$ erg s$^{-1}$ (Hamaguchi et al. 2005; Swartz et al. 2005; Feigelson et al. 2003; Skinner et al. 2004; Stelzer et al. 2006a) but is ~ $10^2$ higher than β Pictoris Moving Group (BPMG) early- to mid-A stars (Hempel et al. 2005). The spectrum is also much softer than typical for strongly accreting Herbig Ae stars and is 5× below the T Tauri companion above 1 keV. It most closely resembles BPMG early F stars such as 51 Eri (Feigelson et al. 2006; Stelzer et al. 2006b). We find $\log(L_x/L_{bol}) = -5.90$, which is also typical of early F stars, suggesting a similar magnetic field configuration. The X-ray luminosity is not indicative of strong mass accretion activity.

#### 3.6.2 FUV Limits

Garcia Lopez et al. (2006) determined accretion rates for 32 Herbig AeBe stars based on net Brγ emission and found that HD 100453A has an accretion rate of $9.12 \times 10^{-9}$ M$_\odot$ yr$^{-1}$. Brγ emission can be due to accretion and to stellar activity and could also be due to the companion. However, Chen et al. (2006) imaged HD 100453 in Brγ and resolved the companion with a measured contrast ratio of 4.2 magnitudes. Thus only 2% of the Brγ flux is contributed by the companion, which is insignificant. If Brγ emission is solely due to accretion, one would expect a high level of FUV continuum. We see a low level of FUV continuum (see § 2.3.1). Simon et al. (2002) have shown that convection occurs in A-stars with $T_{eff} \leq 8250$ K and Hamidouche et al. (2008) suggest that Herbig Ae stars have stellar magnetic activity. Moreover, our FUSE data show excess C III λ1176 Å emission (§ 2.3.1), suggesting stellar activity for HD 100453. We therefore infer that the Brγ emission has an stellar activity component causing the Garcia Lopez et al. (2006) HD 100453A accretion rate estimate to be high.

In Herbig Ae stars, the FUV continuum is the sum of the photospheric emission and mass accretion emission. In § 2.3.1 we have shown that the photospheric continuum emission in the range $1160 \pm 5$ Å is negligible for HD 100453A and DX Cha. Thus, the FUV continuum in this range should be proportional to the accretion luminosity, $L_{acc}$, for each star. From the mean continuum ratio determined in § 2.3.1, we find $\dfrac{L_{acc\,DX\,Cha}}{L_{acc\,HD\,100453}} > 17$. Garcia Lopez et al. (2006) give the relation between accretion luminosity and accretion rate as $\dot{M}_{acc} = L_{acc} R_* G^{-1} M_*^{-1}$. Adopting $T_{eff,\,HD\,100453A}$ = 7390 K, $T_{eff,\,DX\,Cha}$ = 7580 K (Kenyon and Hartmann 1995), $L_{bol,\,HD\,100453A}$ = 9 L$_\odot$ (Dominik et al. 2003), $L_{bol,\,DX\,Cha}$ = 25 L$_\odot$ (Grady et al. 2004), we use the Siess et al. (2000) models to estimate the mass and radius of each star. Taking the ratio and solving, we find $\dfrac{\dot{M}_{acc,\,DX\,Cha}}{\dot{M}_{acc,\,HD\,100453}} > 25$. Garcia Lopez et al. (2006) find a DX Cha accretion rate of $3.5 \times 10^{-8}$ M$_\odot$ yr$^{-1}$, from which we infer a 1σ HD 100453A accretion rate upper limit of $1.4 \times 10^{-9}$ M$_\odot$ yr$^{-1}$ based on the FUSE continuum data.

#### 3.6.3 Lyα Jet and HH Knot Limits

Assuming the gross properties of HH knots around HD 100453 and HD 163296 are similar, we derive a crude estimate on accretion rate by using the fact that Herbig-Haro (HH) knots are Lyα bright (Grady et al. 2004; Devine et al. 2000) for low extinction lines of sight. HD 163296 has HH knot Lyα surface brightness of about $5 \times 10^{-14}$ erg cm$^{-2}$ s$^{-1}$ arcsec$^{-2}$ corresponding to a mass loss estimate of $1 \times 10^{-8}$ M$_\odot$ yr$^{-1}$ (Wassell et al. 2006). We find that the HD 100453A HH knot surface brightness upper limit is ~ 6000× lower than the HD 163296 knot surface brightness, which implies an accretion rate upper limit for HD 100453A of ~$6 \times 10^{-11}$ M$_\odot$ yr$^{-1}$.

#### 3.6.4 Hα Emission

Manoj et al. (2006) surveyed Hα profiles for 91 Herbig B−, A−, and F−type stars, and found HD 100453A to have the weakest observed Hα emission with EW = −0.8 Å. Of the 49 Herbig stars in Acke et al. (2005), HD 100453A has the only Hα profile in absorption, with an EW = 1.5 Å. All other stars in the survey show Hα in emission and most have EWs in the range $10 - 100$ Å. With Hα emission being an indicator of accretion activity, HD 100453A is at best a weak, variable accretor. Our FUSE data show excess C III λ1176 Å emission (§ 2.3.1), suggesting stellar activity for HD 100453. Therefore, we cannot exclude Hα contamination from stellar activity. Table 8 summarizes the accretion rate results.

### 3.7 HD 100453B Accretion

Given the HD 100453B spectral type of M4.0V − M4.5V, we considered the UV spectrum of 2MASSWJ1207334-393254 (2M1207 hereafter) presented by Gizis et al. (2005). 2M1207 is a brown dwarf with a spectral type of M8, mass of 0.03 $M_\odot$ (Gizis 2002), and accretion rate of $\sim 10^{-10}$ $M_\odot$ $yr^{-1}$ (Stelzer et al. 2007). We used the 2M1207 UV emission lines detected by Gizis et al. (2005), scaled to the distance of HD 100453B, as a source in the ACS imaging exposure time calculator (ETC). The ETC shows that 2M1207 would have been a 5σ detection in our SBC F122M imagery, corresponding to a 5σ accretion rate upper limit of $\sim 10^{-10}$ $M_\odot$ $yr^{-1}$ for HD 100453B.

### 3.8 Constraints on Disk Structure

The SED of HD 100453A has strong IR excesses in both the NIR and FIR (M02). However, it has an absence of small, warm dust emission from the disk surface compared to the average for Herbig Ae stars, as evidenced by a lack of the 10 μm silicate feature in its mid-IR spectrum (M01). Note however that large warm silicate grains of a few μm in size may still be present. Dominik et al. (2003) calculate stellar flux $F_*$ from a source's SED by integrating the Kurucz model fit to the optical and UV spectrum. To compute the emission from the disk, they subtract the Kurucz model from the SED and integrate the excess flux from $2 − 7$ μm to define $F_{NIR}$ and from 7 μm to infinity to define $F_{FIR}$. The sources in M01 have a mean $F_{NIR}/F_* = 0.16$ and mean $F_{FIR}/F_* = 0.23$. HD 100453A has $F_{NIR}/F_* = 0.22$ and $F_{FIR}/F_* = 0.29$, numerically demonstrating its higher than average excess in both the NIR and FIR. Such an SED suggests that dust intercepts a significant portion of the primary star's luminosity in both the inner and outer parts of the disk.

#### 3.8.1 The Inner Disk

Carmona et al. (2008) searched for $H_2$ $0 − 0$ S (2) (J = 4 − 2) emission at 12.278 μm and $H_2$ $0 − 0$ S (1) (J = 3 − 1) emission at 17.035 μm and did not detect either toward HD 100453A. Stringent 3σ upper limits to the integrated line fluxes and the mass of optically thin warm gas (T = 150, 300 and 1000 K) in the disk were derived. They determined that the disk of HD 100453A contains less than a few tenths of Jupiter mass of optically thin $H_2$ gas at 150 K, and less than a few Earth masses of optically thin $H_2$ gas at 300 K and higher temperatures.

Brittain et al. (2007) detected warm CO P30 and P31 emission around every Herbig Ae/Be star with an optically thick disk, i.e., $E(K − L) > 1$. Although $E(K − L) = 1.3$ for HD 100453A, we detect no CO emission lines in the Phoenix echelle spectrograph data near 4.97 μm (Fig. 7). Fig. 10a illustrates the relationship between CO luminosity and NIR excess for several sources discussed in Brittain et al. (2007) and HD 100453A (the filled oval). It is clear that HD 100453A is an anomaly.

To determine an upper limit on P30 luminosity, we assumed an unresolved emission line well described by a Gaussian instrument profile with height equal to the standard deviation of the normalized continuum. Based on the M-band magnitude given in Malfait et al. (1998), our derived 2σ upper limit on P30 line luminosity is $1 \times 10^{20}$ W. Fig. 10b illustrates CO luminosity vs. mass accretion rate for several sources (Brittain et al. 2007) and HD 100453A (this work). We see that the HD 100453A data point is consistent with a general relationship between CO luminosity and mass accretion rate.

We see no pumped Fe II emission in the FUSE data (Fig. 3d). Three scenarios can lead to this result. First, if there is high extinction toward HD 100453A, the Fe II emission may not be detected. However, there is very low extinction toward HD 100453A. Second, if there is no Lyα to pump the transitions, we will detect no Fe II emission. However, IUE SWP 53937 shows well-exposed Lyα emission superposed on the image of the large aperture (20″ × 10″) in the light of the geocoronal emission (Fig. 5). Third, no gas in the inner disk of a star will result in an Fe II non-detection (Harper et al. 2001). An Fe II non-detection with low extinction and abundant Lyα to pump suggests no gas in the inner disk of HD 100453A.

#### 3.8.2 The Outer Disk

Inspection of Figure 2c of Kamp et al. (2005) shows that for the disk of a Herbig Ae star, temperatures appropriate to excite CO J = 3−2 emission (50 ± 10 K) exist in the midplane from ~20 to ~225 AU. As discussed in § 4.2, the outer edge of the dust disk of HD 100453A is likely within this range. We detect no CO J = 3−2 (868 μm) emission toward HD 100453A/B. The 1σ upper limit on the integrated intensity in binned 10 km $s^{-1}$ channels is $I_{CO} < 0.077$ K km $s^{-1}$. Compared to the CO survey by Dent et al. (2005), our non-detection, normalized to a distance of 100 pc, corresponds to an integrated intensity, $I_{CO} < 0.1$ K km $s^{-1}$, very similar to the lowest detection in their survey.

The 3σ upper limit on the integrated intensity in binned 5 km $s^{-1}$ channels is $I_{CO} < 0.042$ K km $s^{-1}$. Following Scoville et al. (1986) and assuming an excitation temperature of 50 K, this corresponds to a 3σ upper limit on CO column density of $N_{CO} < 2.0 \times 10^{13}$ $cm^{-2}$, which is well below the optically thick condition of $N_{CO} > 1 \times 10^{15}$ $cm^{-2}$. This implies a 3σ upper limit on the cold gas mass of $< 3.3 \times 10^{-4}$ $M_J$, assuming a CO abundance of $[CO]/[H_2] = 1 \times 10^{-4}$ (as in the ISM), optically thin emission, and allowing for 10% He by number.

Meeus et al. (2003) measure HD 100453 1.2 mm continuum emission of 265 mJy. Dominik et al. (2003) model the SED of HD 100453 and find that an unphysically large disk mass (2 $M_\odot$) is required for the best fit. Assuming a gas-to-dust ratio of 100, Acke et al. (2004) derive a cold disk mass of 22 $M_J$ (cold dust mass of 0.22 $M_J$) following Hillenbrand et al. (1992). We derive a cold dust mass based on the 1.2 mm emission following Andrews & Williams (2005). The conversion to mass depends on the value of the dust opacity per unit mass, $\kappa$. Dust opacity is both a function of dust evolution and chemistry in the disk (Henning & Stognienko 1996) and is not well constrained in our case. For protoplanetary disks, we adopt $\kappa = 2.5$ cm$^2$ g$^{-1}$ at 1.2 mm (scaled from a value at 850 µm in Andrews & Williams 2005) and infer a cold dust mass of 0.08 M$_J$. For debris disks, we adopt $\kappa = 1.2$ cm$^2$ g$^{-1}$ at 1.2 mm (scaled from a value at 850 µm in Zuckerman & Becklin 1993) and infer a cold dust mass of 0.16 M$_J$. The mass is approximately linearly dependent on temperature and we have assumed 50 K in the above. In either case, the gas-to-dust ratio in the cold outer disk (more than ~20 AU from the central star) is no more than $4\times10^{-3}$, assuming optically thin emission and a CO abundance typical of the ISM. See § 4.3 for CO depletion considerations and associated gas-to-dust estimates.

### 3.8.3 Limiting Surface Brightness

Fig. 2c illustrates the psf-subtracted, fully calibrated coronagraph image of HD 100453 in the HST/ACS HRC F606W filter, using a linear intensity scale and a 3 pixel Gaussian kernel smoothing function. All of the observed features in this image within ~3″.0 of the central star are well known instrumental artifacts, whose characteristics and origin are discussed in depth elsewhere (Clampin et al. 2003; Krist et al. 2005). The alternating bands of positive and negative flux which dominate within a radial distance of 2″.25 and extend to a lesser extent out to a radial distance of 3″.0, as well as all narrow radial spikes, are residuals due to incomplete psf-subtraction. The minor extension of these residuals in the E−W direction (position angle of 90° and 270°) is a known imperfection arising from the 3″.0 diameter occulting spot and its occulting finger, which is seen at the eastern edge of the displayed field of view.

To establish the limiting null detection of HD 100453A's scattered light disk in the F606W filter, we extracted the median azimuthally averaged radial profile surface brightness behavior of the image. Outside of the region dominated by psf-subtraction residuals (> 3″), the SNR = 1 limiting surface brightness of the disk is $21.8 - 21.9$ mag arcsec$^{-2}$ (Fig. 2d). As a more conservative estimate, Fig. 2d also illustrates the SNR = 2 limiting surface brightness of the scattered light disk, which is $\sim 21.1 - 21.2$ mag arcsec$^{-2}$ outside of 3″.

The non-detection of the scattered light disk outside the region dominated by psf-subtraction residuals could indicate a complete absence of material at r > 3″, or that dust at r > 3″ isn't being illuminated, or that the grain size distribution or optical depth of the disk is such that little scattered light is produced in the F606W filter band.

## 4. DISCUSSION

### 4.1 The Companion

HD 100453B is a M4.0V − M4.5V star with no detectable NIR excess, little evidence for on-going accretion, an age of $10 \pm 2$ Myr, and $L_x/L_{bol}$ consistent with saturated magnetic activity. The combination of these characteristics suggests it is a weak T Tauri star. Its optical photometry is consistent with that expected for such a star at the distance of HD 100453A. Although we have confirmed its common proper motion with HD 100453A, third epoch observations are required to determine its orbital part about the primary.

### 4.2 The Dust Disk

The inner rim of the disk is estimated to be at less than ~0.2 AU based on the strength of the NIR continuum emission. The trough in the SED of HD 100453A (M02) between the NIR and mid-IR components may indicate a shadowed region or a gap in the disk. The mid-IR spectrum of HD 100453A is rich in polycyclic aromatic hydrocarbons (PAH) features, which have been marginally spatially resolved out to 0″.2 − 0″.3 (Habart et al. 2006), or a projected distance of about $25 − 35$ AU from the central star. The outer disk radius is >25 AU based on the PAH data, the scattered light disk outer radius is <250 AU based on the HST coronagraphic data, and the disk is optically thin (in the optical and NIR) by a projected radius of ~90 AU based on the detection of star C, which is likely a background star. The M-type companion located at a projected distance of 120 AU from the primary star is likely to have tidally truncated the outer edge of the disk. The tightest constraint on the companion's influence occurs if the companion is in the plane of the disk, and if the disk is geometrically thin and is primarily governed by gas pressure and viscosity, but not by self-gravity (Artymowicz & Lubow 1994). With these assumptions, and for the mass ratio of HD 100453A and its companion, Artymowicz & Lubow (1994) predict the outer edge of the disk will be truncated at a projected distance of ~50 AU assuming a circular orbit, and even closer in, if the orbit of the companion is eccentric. A projected disk outer radius of $35 − 50$ AU is in agreement with the spatially resolved PAH emission detected out to a projected radius of $25 − 35$ AU, and the detection of a likely background star at a projected radius of ~90 AU.

### 4.3 The Gas Disk

We have analyzed four accretion rate indicators (FUV continuum, lack of jets/knots, H$\alpha$ emission, and X-ray emission) in this work. From the FUV continuum data, we adopt an accretion rate upper limit of $1.4\times10^{-9}$ M$_\odot$ yr$^{-1}$ (Table 8). In addition, variable, weak H$\alpha$ emission is sometimes present, and may indicate either variable or intermittent low levels of accretion or variable stellar activity.

The non-detection of warm CO and H$_2$ emission toward HD 100453A indicates that there is little optically thin gas in the inner disk (R < 20 AU). Carmona et al. (2008) found that the disk of HD 100453A contains less than a few tenths of Jupiter mass of optically thin H$_2$ gas at 150 K, and less than a few Earth masses of optically thin H$_2$ gas at 300 K and higher temperatures. This has two possible explanations. (1) The first is that little warm gas (CO and H$_2$) and dust exist in the inner disk. The medium is optically thin (because of the low density of dust), but the mass of gas is so low that the emission flux level is not detectable. (2) The second possibility is that there is warm gas residing in the optically thick disk midplane. In this case, any line emission originates from the optically thin molecular surface layer of the disk. However, there is so little gas in the surface layer that the emission level is not detectable (Carmona et al. 2008). Present data do not allow us to distinguish between these two scenarios, but our very low accretion rate upper limit suggests reduced amounts of gas in the inner disk, even for the optically thick case.

Using the most extreme case of CO depletion found by Thi et al. (2001) (a factor of $10^4$), we find a total cold gas mass upper limit of 3.3 $M_J$ which corresponds to a gas-to-dust ratio of no more than ~40:1 in the outer disk. Adopting the more likely case of depletion by a factor $\leq 10^3$ (e.g. Thi et al. 2001), we find a total cold gas mass upper limit of 0.33 $M_J$ which corresponds to a gas-to-dust ratio of no more than ~4:1 in the outer disk. Even for extreme depletion, we find that the outer disk is gas-poor compared to typical protoplanetary disks which are expected to have gas-to-dust ratios of 100:1. More sensitive sub-millimeter observations might reduce the gas:dust ratio upper limit even more.

We do not detect molecular gas emission sufficient to maintain the outer dust disk in a flared geometry (PAHs may be sufficiently small not to settle much, or may be a skin phenomenon on the dust disk), nor do we have evidence for molecular gas in the inner disk, or FUV signatures of accretion in our single epoch FUSE spectrum. Departures from a flared disk geometry are common among transitional disks (Grady et al. 2001 HD 100546; Doering et al. 2007 HD 97048; Grady et al. 2007 HD 169142; Grady et al. 2009 SAO 206462).

### 4.4 Limits on In-disk Objects

Pontoppidan et al. (2008) argue that a low warm gas abundance in the gap or hole of a disk suggests an in-disk stellar companion. Although we find a low gas abundance in the inner disk of HD 100453, our X-ray imagery excludes the presence of any additional co-moving stellar coronal source more luminous than M5.0V in the system. We cannot exclude the presence of a non-accreting brown dwarf, which would exhibit a similar X-ray spectral behavior as HD 100453A (Tsuboi et al. 2003) but be an order of magnitude fainter. While wide star-brown dwarf binaries have been identified in the TW Hya association and BPMG (Feigelson et al. 2006; Stelzer et al. 2006b), none have been identified within 50 AU of a Herbig Ae star.

### 4.5 The Evolutionary State of the Disk

The large fractional IR excess luminosity of HD 100453 has been previously associated with a flared, gas-rich protoplanetary disk (M01, M02). If this is the case, it is unusual that we have detected no hot, warm, or cold molecular gas toward the system. An absence of accretion signatures, the lack of detectable spatially-extended jets, and the non-detection of dissociation products from refractory-rich grains like Fe II suggest HD 100453A is a transitional disk system. Its SED is "doubled-peaked" in the IR and appears to be more representative of a pre-transitional disk system (i.e. a disk with a gap) than a transitional system (i.e. a disk with a hole), though detailed modeling of the system should be done to confirm this interpretation. However, pre-transitional and transitional disks generally have gas in the outer disk and at least some gas in the inner disk. The lack of any detectable molecular gas is more representative of a debris disk. Provided that HD 100453A produces at least $10^{41}$ ionizing photons $s^{-1}$, photoevaporation should result in complete clearing of the disk within $10^5$ yr, once the outer disk mass is reduced to ~1 $M_J$ (Alexander et al. 2006). With an estimated outer disk mass of ~0.2 $M_J$, photoevaporation could be responsible for clearing the primordial disk of HD 100453A. However, if this is the case, an unusually high fractional IR excess luminosity (~$100 - 1000\times$ that of a typical cold debris disk) is resulting from debris disk secondary dust (i.e. dust produced by collisions of massive bodies) in this system. Simple analysis of the Stefan-Boltzmann law shows that a blackbody at T=300 K produces ~$1000\times$ the irradiance of the same blackbody at T=50 K, suggesting a warm debris disk could be capable of producing the level of IR excess luminosity observed. However, detailed modeling of the system should be done to confirm this possibility.

Until the disk is spatially resolved with higher resolution techniques such as interferometry, spectroastrometry, etc., we can only suggest that the disk of HD 100453A is in an unusual evolutionary state between a gas-rich protoplanetary disk and a gas-poor debris disk. Given the low current gas content, gas giant formation, if present, must have ended in this system by $10 \pm 2$ Myr. The low gas content is typical of the ß Pictoris moving group debris disks, which are coeval with HD 100453.

### 5. SUMMARY

In this work, we have used multi-wavelength observations from the X-ray to the millimeter to confirm the binary hypothesis, refine the age of the system, determine the accretion rate, and to constrain the size, gas content, and dust content of the disk environment of HD 100453A.

• We find that HD 100453B is indeed the physical companion to HD 100453A, is separated by a projected distance of about 120 AU, and has a spectral type of M4.0V − M4.5V, a mass of $0.21 - 0.30$ $M_\odot$, and $\log(L_x/L_{bol}) = -3.75$.

• We find that the age of the system is $10 \pm 2$ Myr.

• We find little evidence of accretion or jet activity for HD 100453A. Our FUSE data suggest an accretion rate upper limit of $1.4\times10^{-9}$ $M_\odot$ $yr^{-1}$.

• PAH emission from the disk is marginally resolved out to ~25 − 35 AU (Habart et al. 2006), but we detect a likely background star at ~0″.8 (projected distance of ~90 AU), and we see no evidence of a scattered light disk outside 2″.25 (~250 AU). It is possible that the disk has been tidally truncated by the companion at less than a projected distance of ~50 AU.

• Based on the 1.2 mm continuum flux from the literature, we find that the outer disk of HD 100453A contains ~0.1 $M_J$ of cold dust.

• The combination of *at best* a low accretion rate and the non-detection of several gas tracers suggests that the disk is gas-poor. The non-detection of CO J = 3−2 indicates that the cold gas in the outer disk is optically thin, the outer disk contains no more than ~0.33 $M_J$ of cold molecular gas, and the gas-to-dust ratio in the outer disk is no more than ~4:1 (assuming $[CO]/[H_2]$ abundance of $1 \times 10^{-4}$ and a depletion factor of $10^3$).

To better understand the evolutionary state of the HD 100453 system, interferometry, spectroastrometry, or high resolution, high contrast observations with an occulting disk ≤0″.3 in radius are needed to spatially resolve the disk structure. Additional epochs of proper motion data are required to determine the orbital dynamics of the B companion and to confirm that star C is a background star. The super-high resolution and contrast of a possible future mission such as the Terrestrial Planet Finder might be successful in imaging a large planet in the system.

KAC is supported by a Kentucky Space Grant Consortium Fellowship under NASA National Space Grant College and Fellowship Program Grant NNG05GH07H. This work is performed while K.H. is supported by the NASA Astrobiology Program under CAN 03-OSS-02. JPW is supported by an NSF AAPF under award AST-0802230. Based on observations made with the NASA/ESA *Hubble Space Telescope*, which is operated by the Association of Universities for Research in Astronomy, Inc., under NASA Contract NAS5-26555. Based on observations made with the *Chandra X-Ray Observatory*, which is operated by the Smithsonian Astrophysical Observatory for and on behalf of the National Aeronautics Space Administration under contract NAS8-03060. Based on observations made with the NASA-CNES-CSA *Far Ultraviolet Spectroscopic Explorer*. FUSE is operated for NASA by Johns Hopkins University under NASA contract NAS5-32985. These data were collected under the FUSE GI program C126. Also based on observations obtained at the *Gemini Observatory*, which is operated by the Association of Universities for Research in Astronomy, Inc., under a cooperative agreement with the NSF on behalf of the Gemini partnership: the National Science Foundation (United States), the Particle Physics and Astronomy Research Council (United Kingdom), the National Research Council (Canada), CONICYT (Chile), the Australian Research Council (Australia), CNPq (Brazil) and CONICET (Argentina). The Phoenix infrared spectrograph was developed and is operated by the National Optical Astronomy Observatory. The Phoenix spectra were obtained as part of program GS-2006B-C-3. Based on observations collected at the *European Southern Observatory*, Chile with VLT NACO (program ID 077.C−0570(A)) and with VLT SINFONI (program ID 079.C-0018(A)). The James Clerk Maxwell Telescope is operated by The Joint Astronomy Centre on behalf of the Science and Technology Facilities Council of the United Kingdom, the Netherlands Organization for Scientific Research, and the National Research Council of Canada. Based on archival IUE data collected under observing program ID RM050. A.C. would like to thank M. Janson and M. Gustafsson for their help with SINFONI pipeline products. K.C. would like to thank J. Lauroesch for his help in reducing data. We thank the anonymous referee whose valuable comments have significantly improved the paper.

*Facilities:* HST(ACS), CXO(ACIS-S), FUSE, Gemini:South(Phoenix), VLT:Yepun(NACO, SINFONI), JCMT(HARP), ESO:1.52m(FEROS), IUE

Table 1.

Summary of X-ray Spectral Fit Model Parameters

|  | HD 100453A | HD 100453B | Units |
|---|---|---|---|
| $k$T (90% Confidence Range) | .29 (0.24-0.36) | .70 (0.49-0.84) | keV |
| $F_x^a$ (0.3-2 keV) | 2.7 | 2.3 | $10^{-14}$ ergs cm$^{-2}$ s$^{-1}$ |
| $L_x^a$ (0.3-2 keV) | 4.0 | 3.3 | $10^{28}$ ergs s$^{-1}$ |
| $\chi^2$/d.o.f. (d.o.f.) | 0.22 (4) | 0.43 (5) |  |

[a]divided by 0.81 from the encircled flux within 0″.6 circle

Note: — Elemental abundances are fixed at 0.3 solar.

Table 2.

Summary of HST/ACS Observations

| Object | Date | Instrument | Filter | Mode | Exp. Time (s) | Comment |
|---|---|---|---|---|---|---|
| HD 100453 | 2003 Nov 12 | HRC | F606W | Direct | 20 | |
| HD 100453 | 2003 Nov 12 | HRC | F606W | Coron | 2480 | |
| HD 87427 | 2003 Nov 12 | HRC | F606W | Coron | 2000 | psf-star |
| HD 100453 | 2006 Jun 7 | SBC | F122M | Direct | $4 \times 524$ | SBC−FIX |
| HS2027+0651 | 2002 Jun 09 | SBC | F122M | Direct | 400 | psf-star |
| HS2027+0651 | 2003 Mar 19 | SBC | F122M | Direct | $5 \times 400$ | psf-star |
| HS2027+0651 | 2007 Apr 3 | SBC | F122M | Direct | 402 | psf-star |

Table 3.

Summary of VLT NACO Observations of HD 100453

| Filter | Bandpass (μm) | Date | Mode | Total Exp. Time (s) | Airmass | Seeing |
|---|---|---|---|---|---|---|
| J-band | $1.265 \pm 0.125$ | 2006 May 23 | Coron | 30 | 1.15 | 0″.68 |
| L′-band | $3.80 \pm 0.31$ | 2006 Jun 22 | Coron | 4.32 | 1.17 | 0″.52 |
| M′-band | $4.78 \pm 0.295$ | 2006 Jun 22 | Coron | 7.84 | 1.17 | 0″.52 |
| Brγ | $2.166 \pm 0.012$ | 2006 Jun 22 | Direct | 10 | 1.17 | 0″.52 |

Table 4.

FUV Continuum + Spectral Features

| | $\lambda_{center}$ (Å) | Width (Å) | Mean (ergs cm$^{-2}$ s$^{-1}$ Å$^{-1}$) | Standard Deviation of the Mean (ergs cm$^{-2}$ s$^{-1}$ Å$^{-1}$) |
|---|---|---|---|---|
| DX Cha | 1160.00 | 10.00 | $3.875 \times 10^{-14}$ | $5.260 \times 10^{-16}$ |
| HD 100453 | 1160.00 | 10.00 | $1.936 \times 10^{-15}$ | $3.462 \times 10^{-16}$ |

Table 5.

Summary of VLT SINFONI Observations of HD 100453B

| Filter | Type | # of Offsets | # Exp. × Exp. Time (s) | Total Exp. Time (s) | Jitterbox Size |
|---|---|---|---|---|---|
| K-band | acquisition | 1 | $1 \times 0.83$ | 0.83 | n/a |
| K-band | science | 5 | $24 \times 2$ | 240 | 0″.3 |
| H-band | science | 5 | $18 \times 2$ | 180 | 0″.5 |
| J-band | science | 10 | $12 \times 5$ | 600 | 0″.5 |

Table 6.

Summary of Proper Motion Data

| | HD 100453A | HD 100453B (values *relative to primary*) | Position Angle/ Direction (degrees E of N) |
|---|---|---|---|
| 2003 June 02 Relative Position | – – | 1″.054 ± 0″.007[a] | 127.19 ± 0.30 |
| 2006 June 22 Relative Position | – – | 1″.047 ± 0″.005[b] | 128.22 ± 0.28 |
| 3.05-yr Motion West (mas) | 112.70 ± 2.38[c] | 17.13 ± 5.42 | 270 |
| 3.05-yr Motion South (mas) | 14.40 ± 1.62[c] | 10.54 ± 5.42 | 180 |
| 3.05-yr Motion of B (mas) | – – | 20.11 ± 5.42 | 238.40 ± 28 |
| 3.05-yr Motion Parallel to System Motion (mas) | 113.62 ± 2.37 | 18.33 ± 6.39 | 262.72 ± 0.96 |
| 3.05-yr Motion Orthogonal to System Motion (mas) | – – | 8.28 ± 9.23 | 172.72 ± 0.96 |

[a]Chen et al. (2006); [b]this work; [c]Hipparcos (Perryman et al. 1997)

Table 7.

Summary of Photometry

| Object | Mode | Filter | magnitude | Notes |
|--------|------|--------|-----------|-------|
| HD 100453B | Direct | $m_{F606W}$ | 15.6 | (used in line 3) |
| HD 100453B | Coron | $m_{F606W}$ | 15.8 | (used in line 3) |
| HD 100453B | Combined | $m_{F606W}$ | $15.7 \pm 0.2$ | M4.0V ($\pm$ 1 sub-types) |
| HD 100453B | Direct | $K_s$ | $10.66 \pm 0.1$ | Chen et al. (2006) |
| HD 100453B | Coron | $L'$ | $10.13 \pm 0.1$ | - - - |
| HD 100453B | Coron | $M'$ | $9.99 \pm 0.1$ | - - - |
| HD 100453B | Calculated | V | $15.88 \pm 0.2$ | (see text) |
| HD 100453B | Calculated | K | $10.64 \pm 0.1$ | (see text) |
| HD 100453B | Calculated | L | $10.27 \pm 0.1$ | (see text) |
| HD 100453A | Direct | B | 8.07 | Vieira et al. (2003) |
| HD 100453A | Direct | V | 7.78 | Vieira et al. (2003) |
| HD 100453A Disk | Coron | $m_{F606W}$ | >21.8-21.9 (mag arcsec$^{-2}$) | SNR = 1 detection |
| HD 100453A Disk | Coron | $m_{F606W}$ | >21.1-21.2 (mag arcsec$^{-2}$) | SNR = 2 detection |

Note: — HST and VLT derived photometry for the HD 100453 system. The V, K, and L magnitudes are derived from the $m_{F606W}$, $K_s$, and $L'$ magnitudes, respectively (see text). The HD 100453A disk values correspond to the SNR = 1 and SNR = 2 limiting surface magnitudes of the scattered light disk.

Table 8.

Summary of Accretion Rate Indicators

| Accretion Indicator | Accretion Level | Significance |
|---------------------|-----------------|--------------|
| FUV Continuum | $< 1.4 \times 10^{-9}$ M$_\odot$ yr$^{-1}$ | $1\sigma$ |
| Lack of HH Knots in Ly$\alpha$ | $< \sim 6 \times 10^{-11}$ M$_\odot$ yr$^{-1}$ | factor of 10 |
| H$\alpha$ | weak accretor | – – |
| X-ray | not strong accretor | – – |
| Adopted Accretion Rate Upper Limit | $< 1.4 \times 10^{-9}$ M$_\odot$ yr$^{-1}$ | $1\sigma$ |

# FIGURES

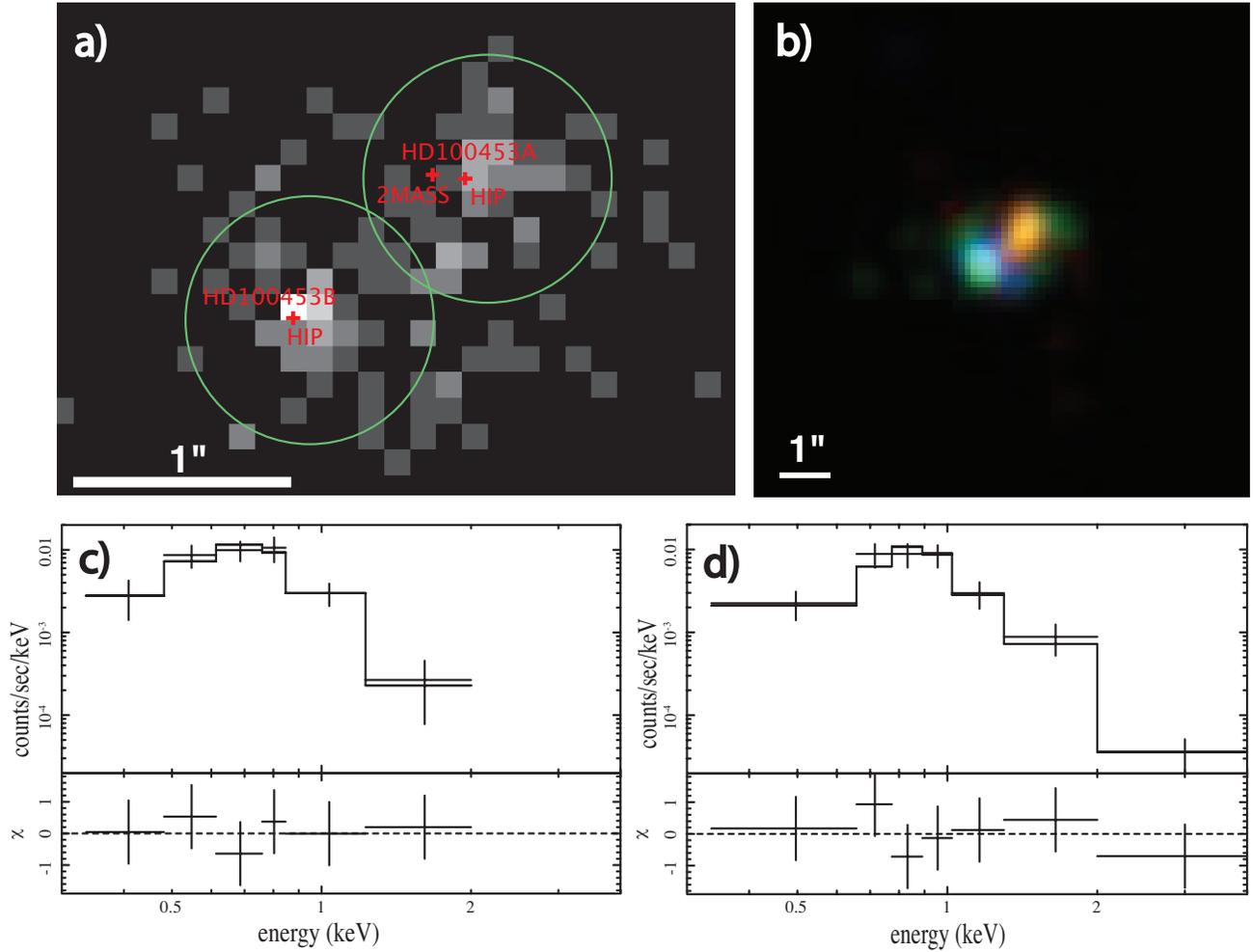

Fig. 1. — a) X-ray image of HD 100453 between $0.35 - 2$ keV. Each pixel has $0''.125$ pixel$^{-1}$. The green circles with $0''.6$ radii show event extraction regions of HD 100453A and HD 100453B. Red crosses show positions of HD 100453A measured with Hipparcos (designated as HIP) and 2MASS (designated as 2MASS). The position of HD 100453B was calculated from the *HIPPARCOS* position of HD 100453A and the separation and position angle in Chen et al. (2006). b) Color-coded X-ray image of HD 100453A/B. The image is smoothed with a Gaussian function with $\sigma = 2$ pixels and color coded in the linear scale to represent $0.35 - 0.7$ keV in red, $0.7 - 0.9$ keV in green and $0.9 - 2$ keV in blue. c) Pulse height spectrum of HD 100453A together with residuals in terms of standard deviations from the single temperature model fit below. The pulse height spectrum is soft, similar to 51 Eri (Feigelson et al. 2006). d) Pulse height spectrum of HD 100453B together with residuals in terms of standard deviations from the single temperature model fit below. The pulse height spectrum is harder than for HD 100453A and is typical of a T Tauri star.

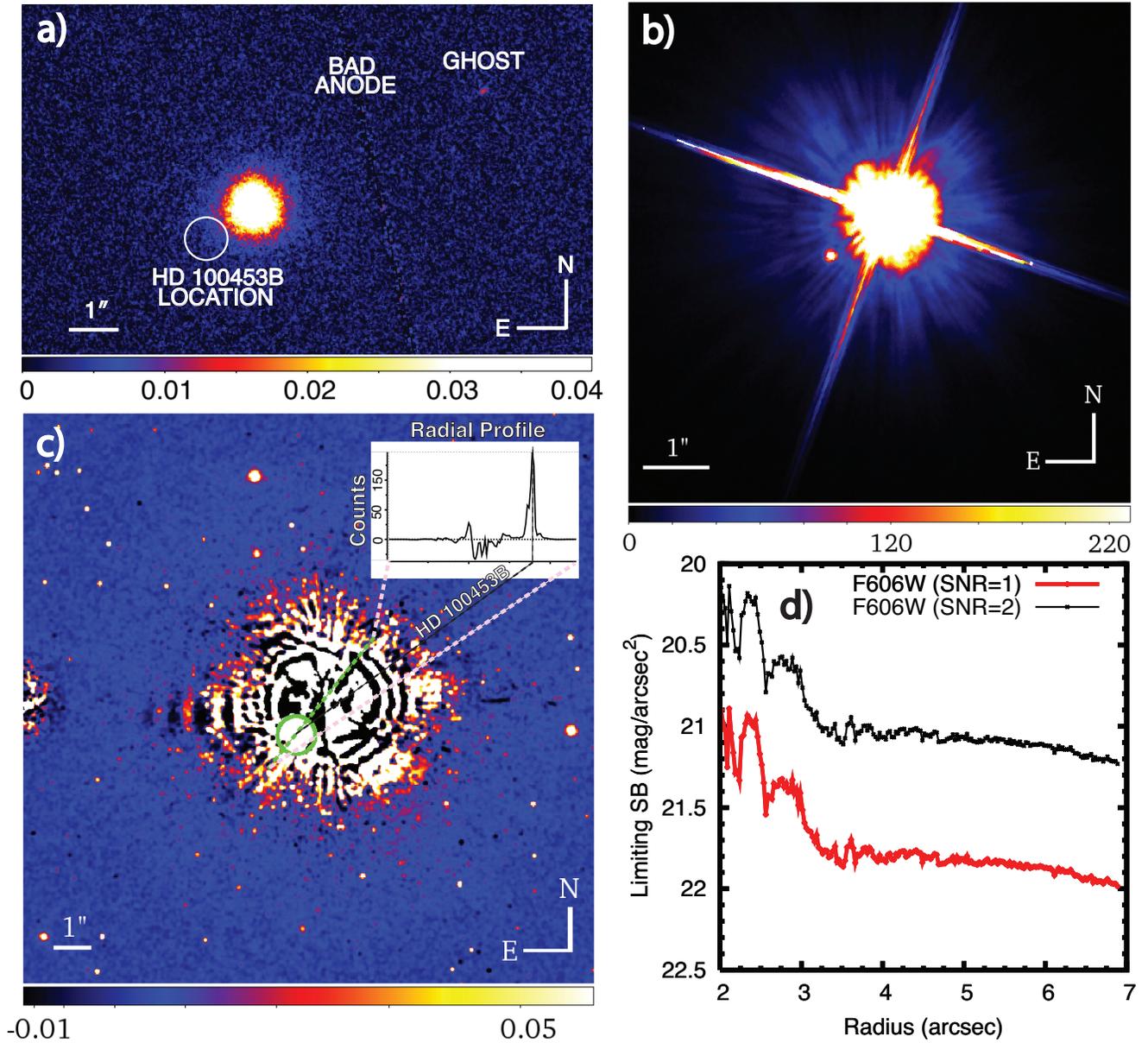

Fig. 2. — HST Imagery and Limiting Surface Brightness. a) A 9″ × 5″.5 HST/ACS SBC direct image of HD 100453 in the F122M filter, plotted on a linear scale. The circle identifies the expected position HD 100453B. No evidence of HD 100453B is detected in the F122M FUV filter, suggesting that it is non-accreting. b) A 7″.5 × 7″.5 direct F606W image of HD 100453A taken with the HST/ACS HRC camera. While the central HD 100453A source is clearly saturated, a distinct second source is clearly visible ~1″.05 southeast from HD 100453A, at a position angle of 126°. c) A 15″.0 × 15″.0 psf-subtracted coronagraphic image of HD 100453 in the F606W filter, plotted on a linear scale and smoothed using a 3 × 3 Gaussian kernel. No evidence of an extended scattered light disk is seen at a level of ~21.2 mag arcsec⁻². HD 100453B, identified within the green circle, is clearly detected through the psf-subtraction residuals (see inset radial profile) and consistent with the position determined from the direct imaging. The green dashed line indicates the location of the radial profile cut displayed in the inset figure. d) The limiting surface brightness of HD 100453A's scattered light disk. We have a null detection of the scattered light disk in our F606W data. Both the SNR = 1 (red) and SNR = 2 (black) limiting surface brightness detection levels were computed from median azimuthally averaged radial profiles of the system.

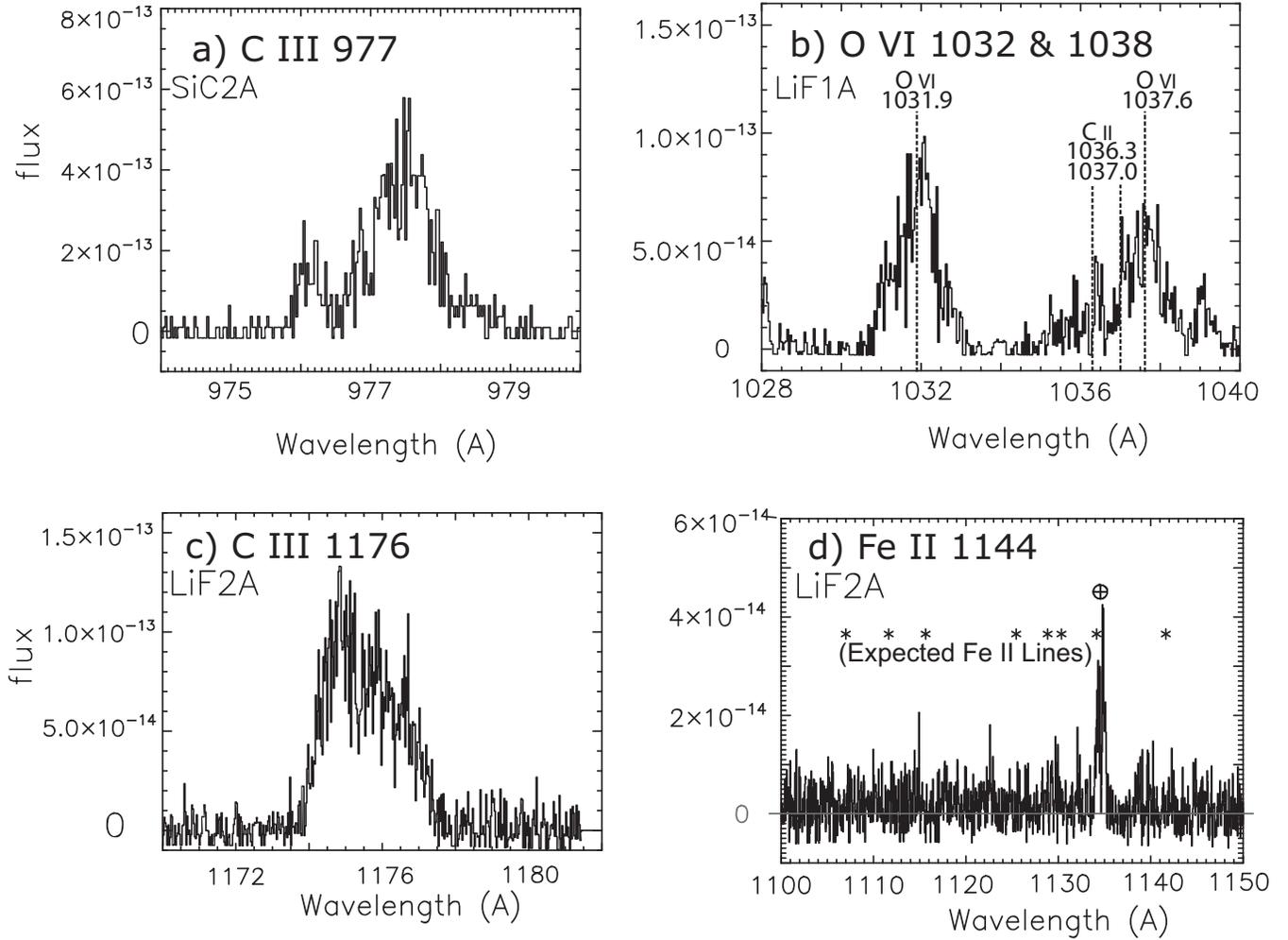

Fig. 3. — FUSE spectra of HD 100453 with flux in units of erg cm$^{-2}$ s$^{-1}$ Å$^{-1}$. a) The C III λ977 Å emission profile has an absorption feature centered near 976.4 Å which is likely O I λ976.4 Å absorption. The detection of C III λ977 at 111 pc suggests low extinction toward HD 100453. b) The flux in the O VI λ1031.9 Å emission profile is 8.8×10$^{-14}$ ± 3.2×10$^{-15}$ erg cm$^{-2}$ s$^{-1}$ Å$^{-1}$. The O VI λ1037.6 Å profile is contaminated with C II λ1036.3 Å and C II λ1037.0 Å. c) The flux in the C III λ1176 Å emission profile is 2.1×10$^{-13}$ ± 3.7×10$^{-15}$ erg cm$^{-2}$ s$^{-1}$ Å$^{-1}$. d) We find no evidence of Fe II λ1144 Å emission, suggesting no gas in the inner disk.

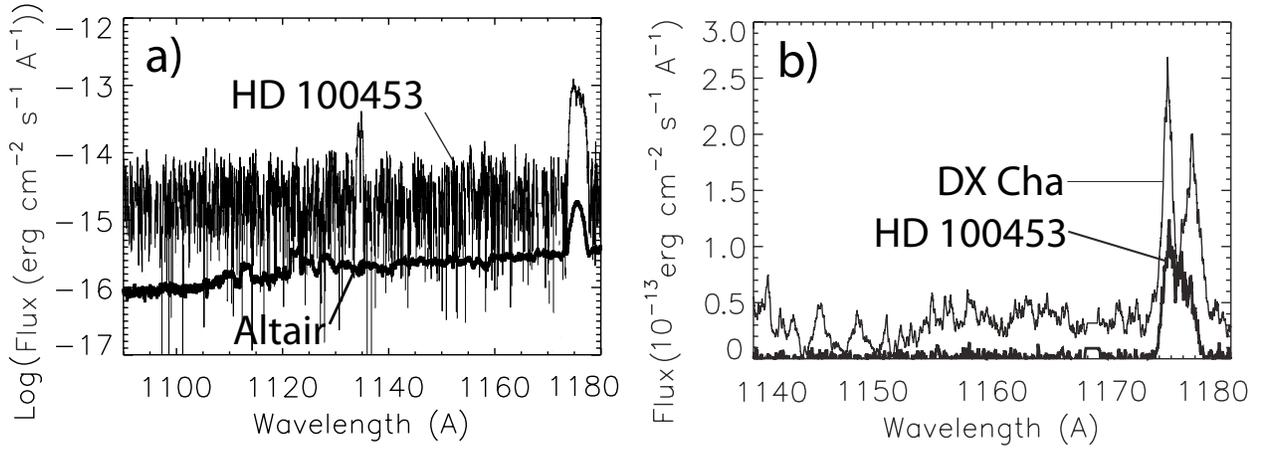

Fig. 4. — FUSE continuum and C III λ1176 Å emission. All spectra have been smoothed by a 9 point running boxcar filter. a) When compared to Altair scaled to the same V magnitude as HD 100453, HD 100453 has excess light in C III λ1176 Å. The photosphere of an A7V star lies far below the limiting flux of $1.9 \times 10^{-15}$ erg cm$^{-2}$ s$^{-1}$ Å$^{-1}$ in the HD 100453 spectrum. We therefore neglect any photospheric contribution to the FUSE spectrum of HD 100453. b) When compared to DX Cha, which is located at essentially the same distance as HD 100453 and with a similar lack of foreground extinction, the upper limit to the FUV continuum light near 1160 Å is a factor of 17 below the continuum detected in DX Cha. Geocoronal emission is masked out in the spectra.

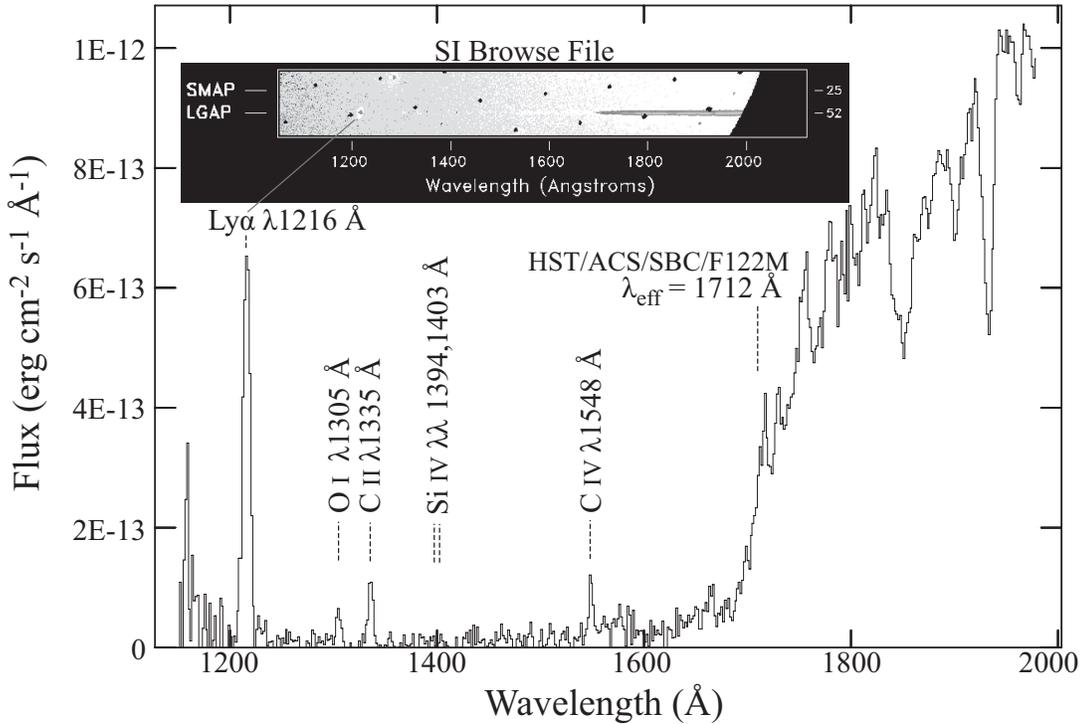

Fig. 5. — The IUE SWP 53937 far-UV spectrum of HD 100453A. Lyα is in emission. The inset figure shows the SI browse file from MAST indicating point source Lyα in emission superposed on the more nebulous geocoronal emission filling the IUE large aperture (LGAP) (20″ × 10″ oval). Also indicated is the effective wavelength of the HST/ACS/SBC F122M filter. Red leakage in the filter causes $\lambda_{eff}$ to shift from the filter center wavelength of 1210 Å to $\lambda_{eff} = 1712$ Å.

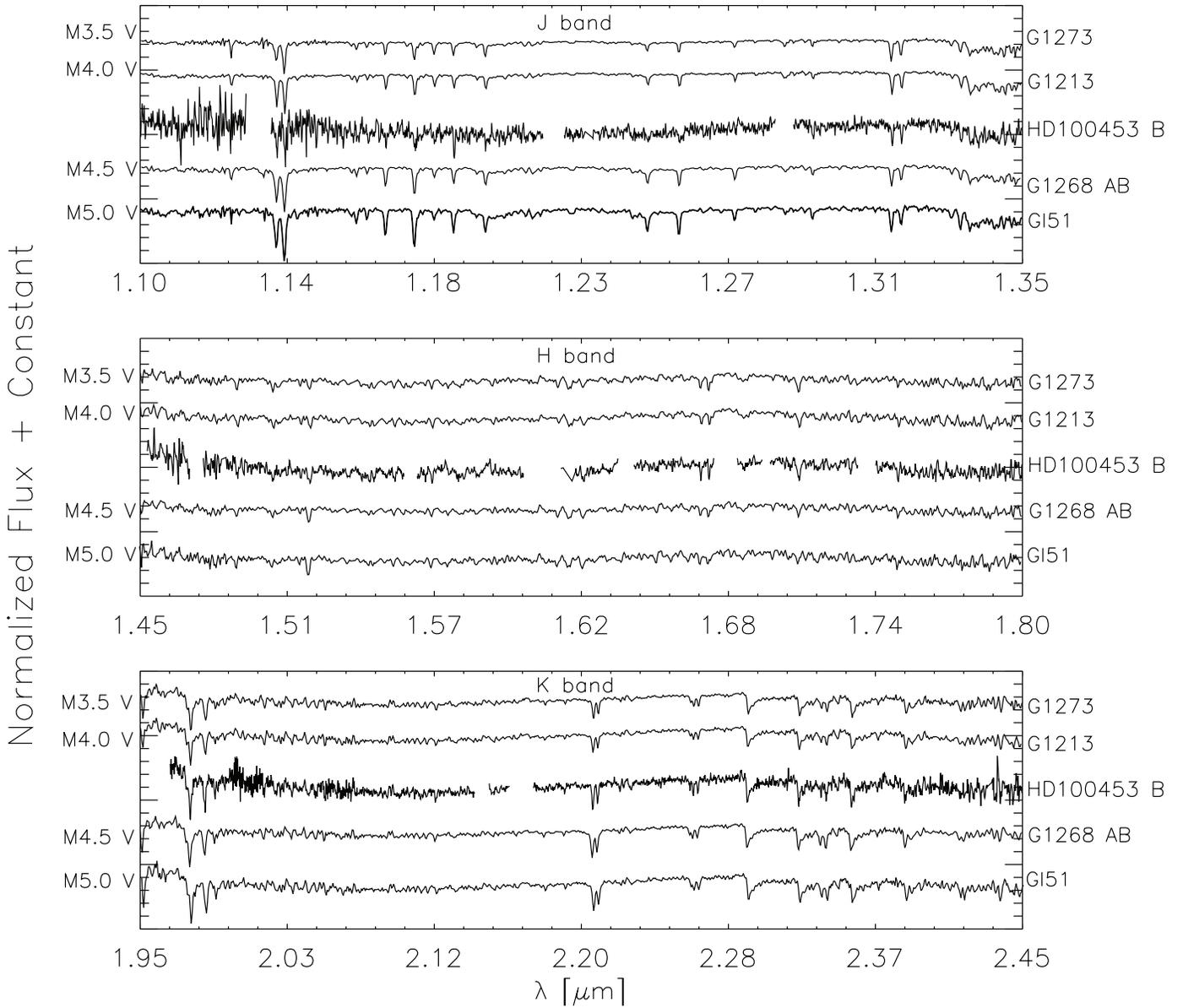

Fig. 6.— SINFONI spectra of HD 100453B in the J, H and K bands. Regions of poor telluric correction (i.e., strong telluric absorption), and regions in which the spectra of the standard star have spectral features have been removed from the spectra. The spectra of four templates from the IRTF NIR library are overplotted in each panel. The spectra are continuum-normalized by a second-degree polynomial.

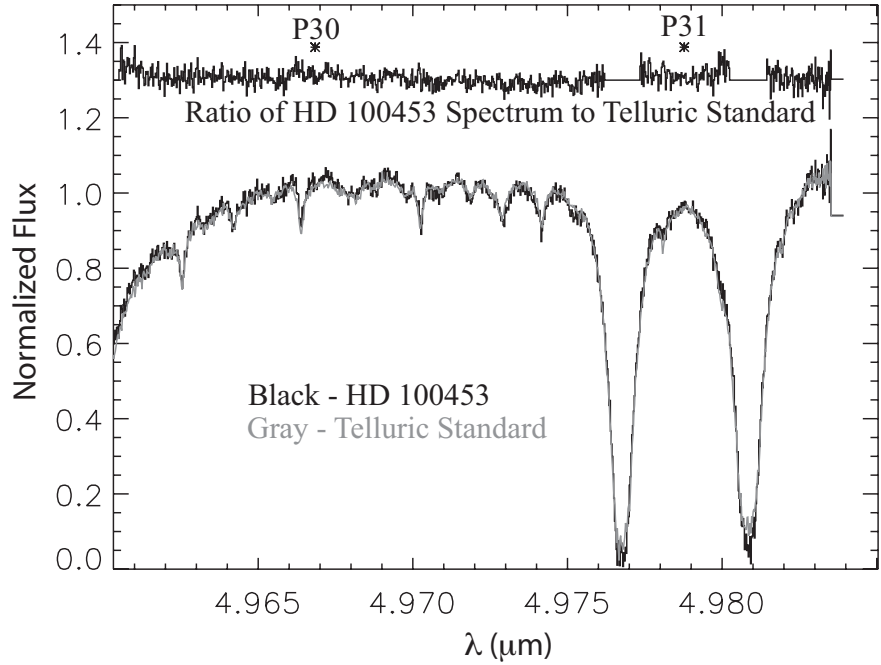

Fig. 7. — Phoenix echelle M-band spectra of HD 100453. The lower black curve traces the HD 100453 spectrum. The lower gray curve traces the telluric standard HR 5671 spectrum. The upper black line is the HD 100453 spectrum corrected for telluric absorption. The gaps in the spectrum are of areas with less than 50% transmittance. The asterisks labeled P30 and P31 mark the rest positions of the CO P30 and P31 rotation-vibration transition lines, which were not detected above noise in our observations.

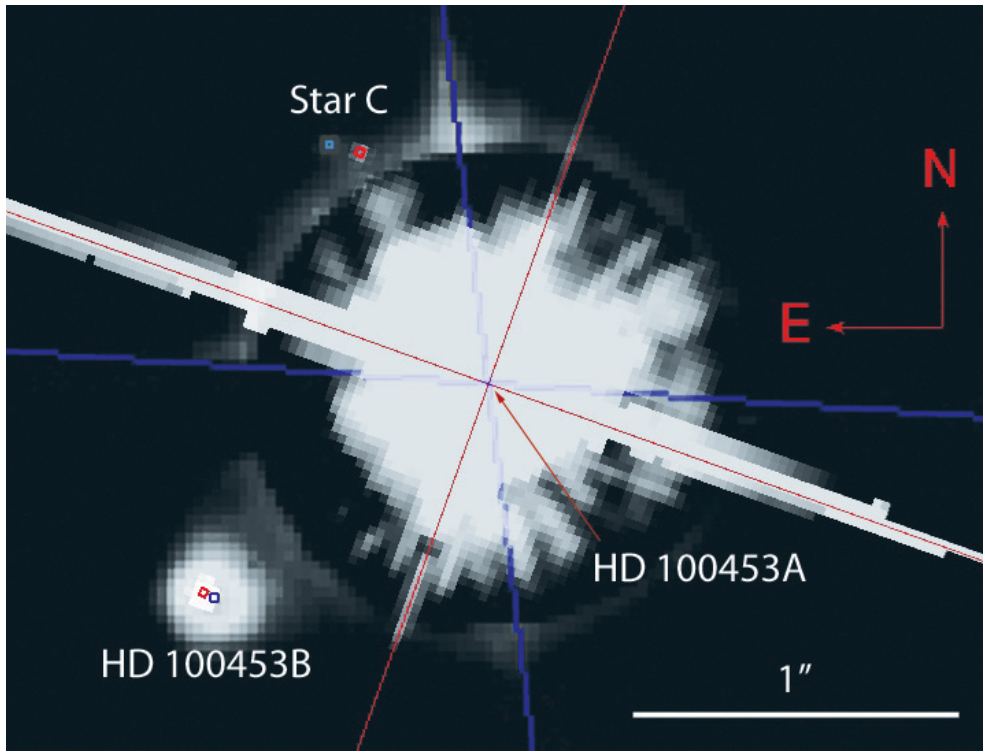

Fig. 8. — Relative proper motion for star C. The image shows an overlay of the 2006 May NACO image onto the 2003 November ACS image with both images oriented such that north is up and east is to the left. For clarity, the center of the ACS and NACO diffraction spikes and object locations have been marked in red and blue, respectively.

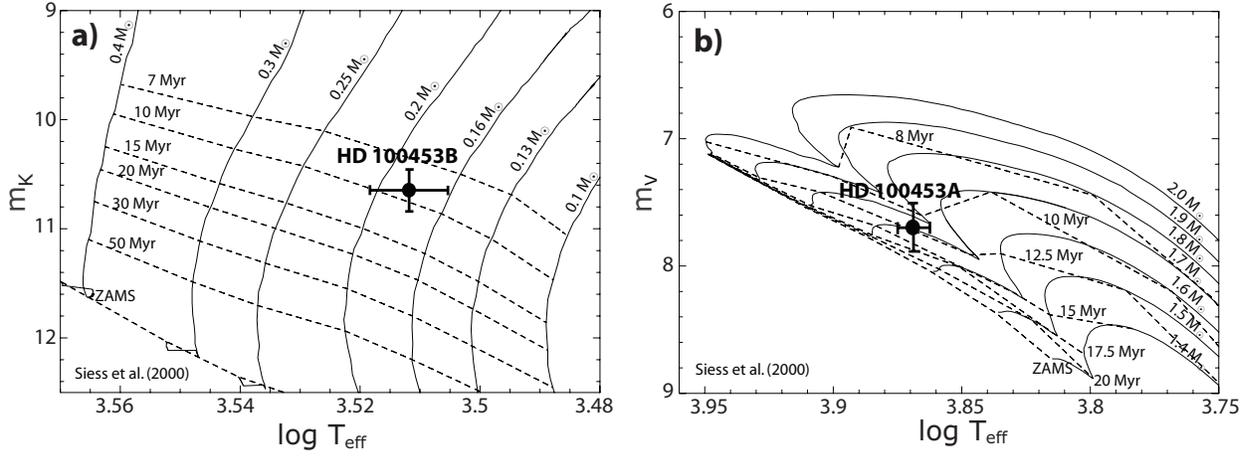

Fig. 9. — a) H–R diagram for HD 100453B. The solid cross locates HD 100453B in the diagram. Based on the Siess et al. (2000) models, we find the age of the system to be $8 - 12$ Myr and the mass is $0.16 - 0.21$ $M_\odot$. b) H–R diagram for HD 100453A. The solid cross locates HD 100453A in the diagram. We find that the age of HD 100453A is $9 - 18$ Myr and the mass is $1.65 - 1.82$ $M_\odot$.

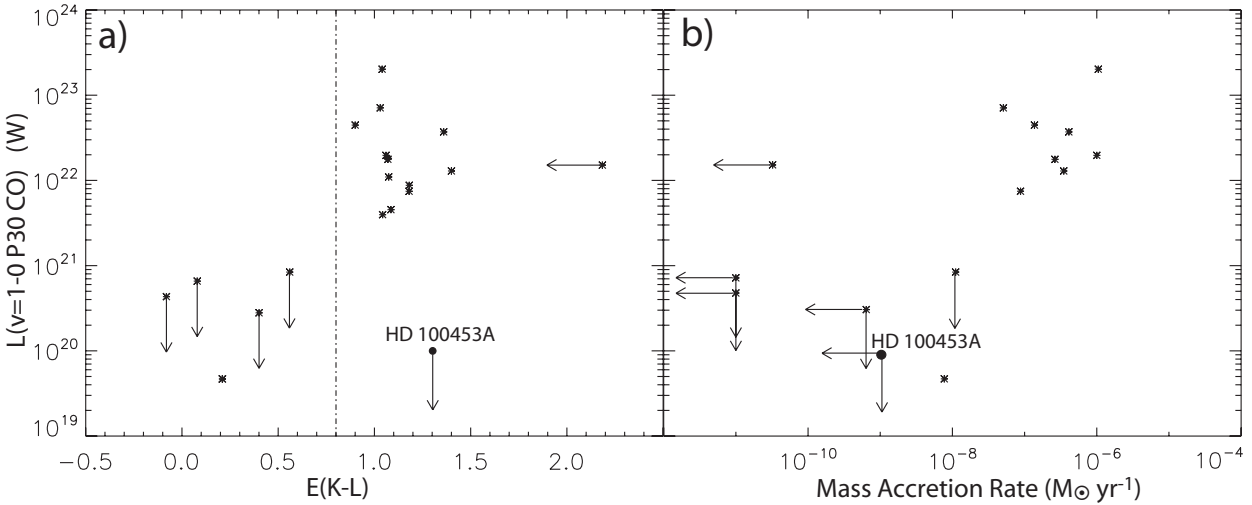

Fig. 10. — a) Herbig Ae/Be star CO luminosity vs. NIR excess (adapted from Brittain et al. 2007). All of the Herbig Ae/Be stars with an optically thick disk (i.e. $E(K - L) > 1$) have detected CO emission, except HD 100453A which is indicated by the filled circle. We find a $2\sigma$ upper limit for the luminosity of HD 100453 of $1 \times 10^{20}$ W, based on the noise floor of the spectrum. A level of $E(K - L) = 1.3$ indicates hot dust in the system and no CO emission suggests reduced levels of gas within $\sim 1$ AU. b) Herbig Ae/Be star CO luminosity vs. mass accretion rate (adapted from Brittain et al. 2007). The luminosity and accretion rate data were taken from Brittain et al. (2007) except for HD 100453A (this work) which is indicated by the filled oval. We see that the HD 100453A data point is consistent with a general relationship between CO luminosity and mass accretion rate.